\documentstyle[sprocl,epsf]{article}
\def\be{\begin{equation}}
\def\ee{\end{equation}}
\def\bea{\begin{eqnarray}}
\def\eea{\end{eqnarray}}
\def\ApJ{\em Ap.~J.}
\def\PRL{\em Phys.~Rev.~Lett.}
\def\PRD{{\em Phys.~Rev.}~D}
\def\Journal#1#2#3#4{{#1} {\bf #2}, #3 (#4)}
\begin{document}
\title{BINARY NEUTRON STARS IN QUASI-EQUILIBRIUM CIRCULAR ORBIT:
	A FULLY RELATIVISTIC TREATMENT}
\author{T.~W.~BAUMGARTE, S.~L.~SHAPIRO}
\address{Department of Physics. University of Illinois at
	Urbana-Champaign, Urbana, Il~61801}
\author{G.~B.~COOK, M.~A.~SCHEEL and S.~A.~TEUKOLSKY}
\address{Center for Radiophysics and Space Research, Cornell University,
	Ithaca, NY 14853}
\maketitle
\abstracts{We present a numerical scheme that solves the initial value problem
in full general relativity for a binary neutron star in quasi-equilibrium.  
While Newtonian gravity allows for a strict equilibrium, a relativistic 
binary system emits gravitational radiation, causing the system to lose 
energy and slowly spiral inwards.  However, since inspiral occurs on a 
time scale much longer than the orbital period, we can adopt a 
quasi-equilibrium approximation. In this approximation, we integrate a 
subset of the Einstein equations coupled to the equations of relativistic 
hydrodynamics to solve the initial value problem for binaries of arbitrary 
separation, down to the innermost stable orbit.}

Neutron star binaries are of interest for several reasons.
They exist, even within our own galaxy, and are among the most 
promising sources for gravitational wave detectors like LIGO, VIRGO and GEO.
More fundamentally the two-body problem is
one of the outstanding unsolved problems in classical general relativity.

So far most researchers have treated binary neutron stars in Newtonian 
theory~\cite{newton}. In post-Newtonian treatments~\cite{postnewton}
the stars are usually treated as point-sources, so 
that hydrodynamical effects are absent. More recently,
Nakamura~\cite{n94} and Wilson and Mathews~\cite{wm95} have initiated
studies of binary neutron stars in general relativity.

In our work we assume that the two stars have 
equal mass, are co-rotating and obey a 
polytropic equation of state, $P = K \rho_0^{\Gamma}$. In Newtonian
gravity, a strict equilibrium solution for two stars in circular
orbit can be found. Since this solution is stationary, the 
hydrodynamical equations reduce to the Bernoulli equation,
\be \label{bernoulli}
\frac{\Gamma}{\Gamma-1} \frac{P}{\rho_0} + 
	\Phi - \frac{1}{2} \Omega^2 (x^2 + y^2) = C,
\ee
where $C$ is a constant, $\Phi$ the gravitational
potential, $\Omega$ the angular velocity, and where the 
rotation is about the $z$-axis. The gravitational potential satisfies
Poisson's equation, $\nabla^2 \Phi = 4 \pi \rho_0$.
These equations comprise a coupled system, containing a
linear elliptic PDE in 3D, which must be solved numerically.

\setlength{\unitlength}{1in}
\begin{picture}(4.2,3.3) 
\put(0,1.3){\epsfxsize=2in \epsffile{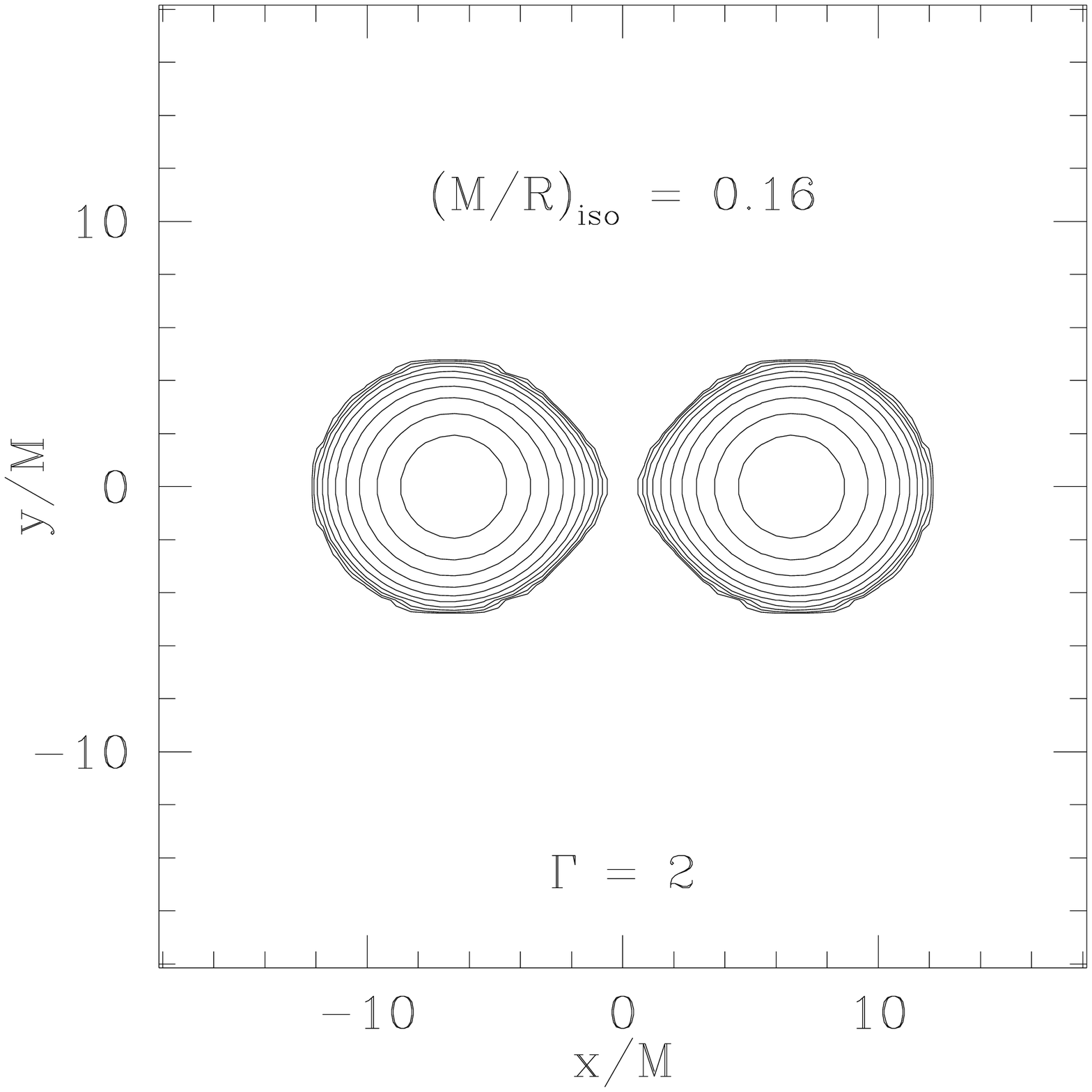} }
\put(2.2,1.3){\epsfxsize=2in \epsffile{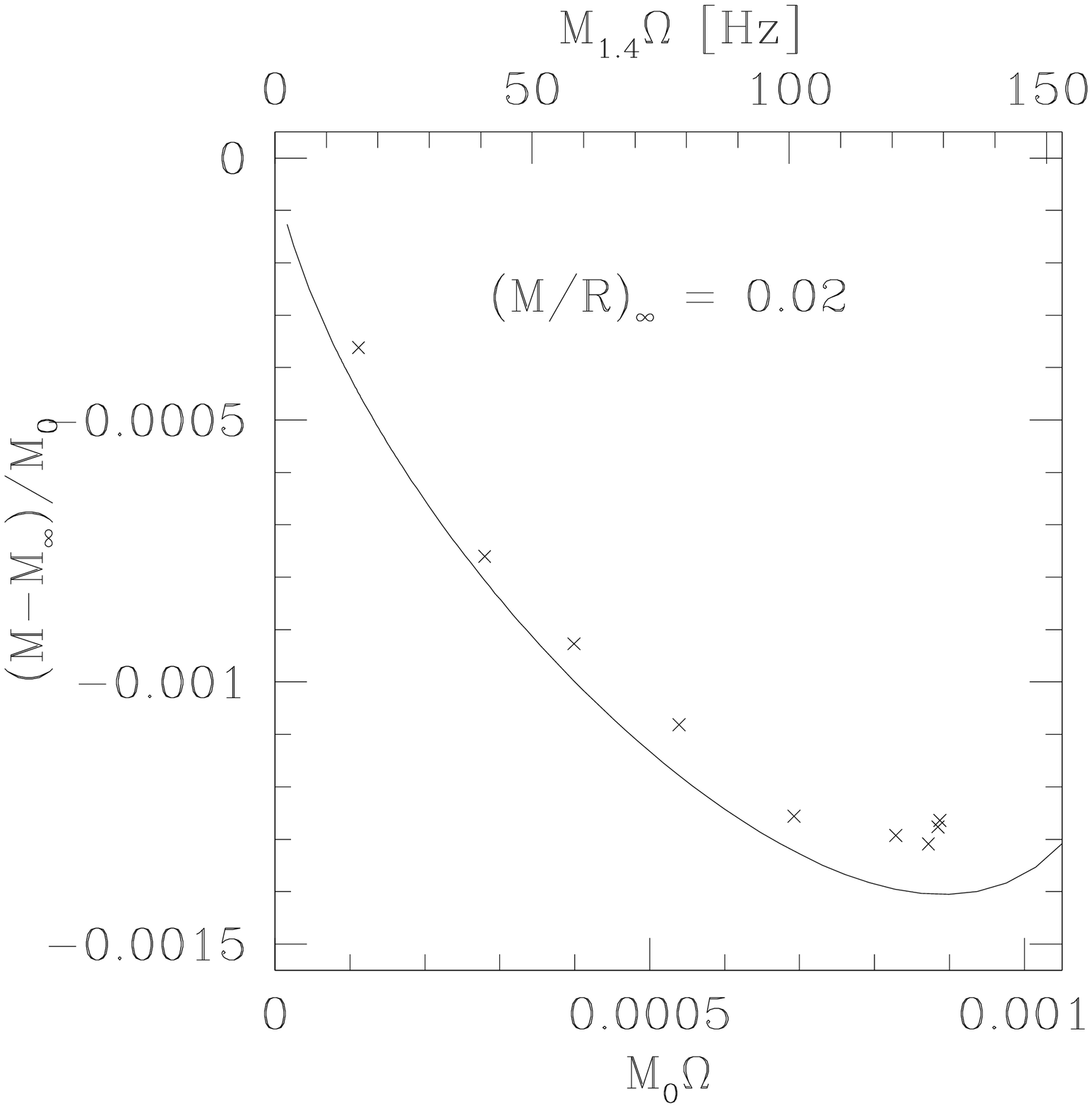} }
\put(0,1.1){\parbox[t]{2in}{Figure 1: Rest 
density contours of a close, 
highly relativistic neutron star binary in the equatorial plane. 
The contours span densities logarithmically between the
central density and 1 percent of that value.}}
\put(2.2,1.1){\parbox[t]{2in}{Figure 2: Binding energy of a $\Gamma = 2$ 
polytrope. The solid line is a result of a post-Newtonian calculation for 
compressible ellipsoids~\cite{lrs97}; crosses are the result of this work.}}
\end{picture}

Because of the emission of gravitational waves, a binary in general 
relativity cannot be in strict equilibrium. However,
up to the innermost stable circular orbit (ISCO) the timescale for 
orbital decay by radiation 
will be much longer than the orbital period so that 
the binary can be considered to be in ``quasi-equilibrium''. This 
allows us to neglect both gravitational waves and wave-induced 
deviations from a circular orbit to good approximation.

To minimize the gravitational wave content we choose the 3-metric to be 
conformally flat~\cite{wm89}. The field equations then
reduce to a set of coupled, quasi-linear elliptic equations for the lapse,
shift, and the conformal factor. Neglecting small deviations from circular
orbit, the fluid flow is again stationary, and the hydrodynamical equations 
reduce to the relativistic Bernoulli equation.
Solving these equations yields a valid solution to the initial value 
(constraint) equations and an approximate solution to the full Einstein 
equations at any given moment, prior to plunge.

As in the Newtonian case, a system of coupled elliptic equations must
be solved, and since the two problems have a very similar structure,
they can both be solved with very similar numerical methods. 
We have developed parallel FAS multigrid solvers for both applications. 
Because of the symmetries of the problem, is is sufficient to work
in one octant. The codes are written in cartesian coordinates and use the 
DAGH infrastructure that has been developed as part of the Binary Black
Hole Grand Challenge project. For code development we typically 
run on 8 processors on the IBM SP2 parallel cluster at Cornell. We
use up to 5 levels of refinement, which gives a 
$(64)^3$ grid on the finest level. The matter is covered by about 20
gridpoints in each direction.

We are interested in quasi-equilibrium models in their own right, 
but we also plan to use the models as initial data for fully
relativistic evolution codes. We show a density profile
of a close neutron star binary in Figure~1. 
In isolation, each star would have a compaction of
$(M/R)_{\infty} = 0.16$, showing that this configuration is highly 
relativistic. The maximum mass configuration for this $\Gamma$ satisfies
$(M/R)_{\infty} = 0.22$.

We construct quasi-equilibrium sequences for binaries of fixed
rest mass. Up to the ISCO, these sequences approximate evolutionary
sequences. We plot in Figure~2 
the binding energy $(M-M_{\infty})/M$ versus the separation, 
parameterized by the angular velocity, for a mildly relativistic
sequence ($(M/R)_{\infty} = 0.02$). The turning point of this curve 
indicates the onset of orbital instability at the
ISCO and the angular velocity there. Note that we are restricted
to co-rotating sequences. Sequences of conserved circulation are probably 
more realistic, since maintaining co-rotation would require excessive 
viscosity~\cite{vis}.

In the near future we plan to implement adaptive mesh refinement (AMR)
and increase the accuracy of our calculation. We will then explore the
physics of fully relativistic binary neutron stars for different polytropic
indices, separations, and values of $(M/R)_{\infty}$.

\end{document}